\documentclass[11pt]{article}
  \usepackage{amssymb}
  \usepackage{graphicx}
   \usepackage{latexsym}
    \usepackage{mathrsfs}
     \usepackage{feynmf}
       \usepackage{balance}
    \usepackage{amsfonts}
      \usepackage{color}
\usepackage[round]{natbib}

\title{Magnetohydrodynamic equations for cold quark gluon plasmas: Multi fluidity and Solitary wave stability}
\author{Azam Ghaani\thanks{Email: az.ghaani@stu-mail.um.ac.ir} and 
Kurosh Javidan\thanks{Email: Javidan@um.ac.ir} \\ 
\small{Department of Physics, Ferdowsi University of Mashhad}\\
\small{91775-1436  Mashhad, Iran} }
\date{}
\begin{document}
\maketitle
\begin{abstract}
By means of magnetohydrodynamic equations in a non relativistic multi fluid framework, we study the behavior of  small amplitude perturbations in cold Quark Gluon Plasmas (QGP). Magnetohydrodynamic equations, along with the QGP equation of state are expanded using the reductive perturbation method. It is shown that such a medium should be considered as multi fluid magnetohydrodynamic (MHD) system. The result is a nonlinear wave equation which complies with a modified form of the "derivative nonlinear Schrodinger" equation instead of the KdV equation. We show that the complete set of equations, by considering the magnetic field which is supported by the Maxwell's equations, create stable solitary waves. An interesting result is the existence of an electric field component along the direction of magnetic field which causes charge separability in the medium. Properties of this solitonic solution is studied by considering different values for the QGP characters such as background mass density and strength of the magnetic field (at the scale of compact stars).
\end{abstract}
{\bf{Keywords}}: quark gluon plasma , QGP equation of state , magnetohydrodynamics (MHD) ,  reductive perturbation method , Solitary waves.\\  

\section*{1. Introduction}
Soon after establishing the idea of asymptotic freedom in quantum chromodynamics (QCD), possibility of the existence of quark gluon plasma (QGP) or quark matter at high temperature (above $150 \sim 200 MeV$) and/or high density (upper than the nuclear density, $\rho_{0}\sim3.0 \times 10^{17} Kg/m^3$) was established \citep{1,2,3,4}. Based on these two conditions, we should expect to find QGP in few milliseconds after the Big Bang \citep{3,5}, at initial states of high energy heavy ion collisions which is currently being experimentally pursued \citep{6,3}, and in the core of super dense astronomical objects, such as neutron stars, magnetars and quark stars \citep{2,7,3,8}.

At sufficiently high densities and low temperatures, as in the dense interior of massive neutron stars, hadrons melt into cold quark matter consist of a Fermi sea of free quarks \citep{9,10}. Nowadays the astronomical observations indicate  the existence of huge magnetic field, from $ 10^{8} T $ at the surface to $ 10^{13} T $ in the core of neutron stars \citep{12,13,14}. Also the collapse of a white dwarf to a neutron star happen with an extremely strong magnetic field \citep{15}. Therefore, it is not surprising that a large number of recent works have been presented to investigate the role of strong magnetic fields on the behaviour of dense quark matter.\

But there are few studies on the collective behavior and long range interactions in the quark matter and QGP, modified by magnetic fields in the framework of magnetohydrodynamics (MHD). The set of magnetohydrodynamic equations are a combination of: I) fluid dynamic equations which the Navier-Stokes equation is essentially the simplest equation describing the motion of a fluid  in the non-relativistic framework \citep{16}, II) the continuity equations and III) the Maxwell's equations. These differential equations with different types of nonlinearity and dispersion terms, mainly can be solved numerically. Important characters of propagating waves inside these media are derived by using an approximation method which is able to save the nonlinear behaviours of equations and provide small amplitude solutions describing long range effects in the system. This is the "Reductive Perturbation Method" (RPM) which is a helpful technique preserving nonlinear, dispersive and dissipative effects of the QGP medium in relevant differential equations \citep{p1, rpm 1}. \

A number of striking works on the propagation of nonlinear waves inside the QGP, mostly in the absence of magnetic field through the continuity and momentum equations and considering different models for the equation of state, have been presented which predict the existence of unstable long range behaviours in the framework of the KdV like equations with breaking and shock profiles \citep{ Non-linear, rpm 1, viscose}.  Almost all presented equations of state, containing special parts which provide a nonlinear term in the equation of motion. But for establishing stable solitonic solution in the framework of KdV equation we need a dispersion term too. This part of the KdV equation can be created through a Laplacian term in the energy density which adds a cubic derivative respect to the space coordinate. In the electromagnetic plasmas, this issue is provided by the Laplace equation of the electric potential, while in the QGP there is not same situation at the first order of approximation.  Indeed the most accepted field theoretical models do not have higher derivative terms in their leading orders. Such these terms may be appeared in higher orders of approximations which are usually negligible \citep{fogaca2007}. One can expand the mass density according to the derivatives of space coordinate, if the mass density of the medium is not in its equilibrium state \citep{Fowler}. Therefore higher derivatives are able to create the dispersive term in the KdV equation. \ 

The gluon field in cold quark gluon plasmas is very large, because of high densities in such these situations. If we assume that the coupling constant is not very small and is not spatially constant, the existence of intense gluon field implies that the bosonic fields tend to have large occupation numbers, and therefore higher derivative terms is appeared in the spatial expansion of related term in the energy density \citep{fogaca2011}. Using above models (as well as other methods) one can add a weak dispersive term to the KdV equation, however it is sufficient for stabilizing small amplitude solitary waves. We have shown that, such localized effects, even in the shape of breaking waves, are long lasting enough to create detectable signatures in the border of QGP medium \citep{Raf}.\

Most of the QGP media have been created in an electromagnetically reach environments. Neutron stars, pulsars and magnetars are examples of compact astrophysical objects where the cold and dense nuclear matter and/or  QGP may exist. The magnetic field in such objects typically is very high. This means that investigating behaviour of these media without considering the magnetic field can not be realistic. It is shown that presenting a fixed extreme but external magnetic field stabilizes breaking or shock waves in cold QGP media. In other words, if external magnetic field is enough strong, breaking waves change into stable solitary profiles which are governed by the Zakharov–Kuznetsov (ZK) equation \citep{RA} (but not a KdV equation!). In such situations we have not need to consider small nonlinear terms which are generally are so weak. Indeed the solitonic solutions of the ZK model are stable by themselves. A realistic result can be extracted through solving the full equations of the system by considering the internal magnetic field created in the QGP matter by itself. According to our best of knowledge, evolution of localized waves due to collective behaviours and long rage interactions by considering the full equations of the system have not been investigated before. It is clear that without these results, our knowledge about the wave propagation in such media is incomplete. We will see that the results are completely different with our previous knowledge. Here we focus on models in which the quark matter or the  QGP environment consist of two light flavors (u and d quarks). The bulk QGP is given as a near ideal Fermi liquid and the MIT bag model is used as equaion of state. It may be noted that, this model is not able to create any dispersive term. We have used the the Fermi-Dirac distribution for quarks and the Bose-Einstein distribution for gluons in the medium \citep{rpm 1, RA}. Cold QGP in an external magnetic field as a multi fluid system has been investigated recently, but without considering the self magnetic field due to QGP constituents \citep{Mag}. Stability conditions have been investigated in this paper, but propagation of nonlinear waves has not been studied. Indeed we have to consider induced magnetic field due to motion of plasma particles as well as nonlinear wave propagation to find more realistic knowledge about the system and collective effects therein.  \

Outlines of this paper are as follows: In the next section we briefly present the QGP equation of state according to the MIT bag model. In the section III a review for the magnetohydrodynamic equations in non-relativistic QGP is given. We expand  the system of equations using the reductive perturbation method and derive a nonlinear equation for system variables at zero temperature, in the section IV. The solitonic solution for the derived equation is obtained in section V. In the section VI we discuss on the properties of the localized wave describing the evolution of transverse magnetic field perturbations and the last section is devoted to some concluding remarks.

\section*{2. Magnetohydrodynamic equations of QGP: Multi fluidity}

QGP species have different charges and due to the magnetic field, they find different trajectories in their media. This means that, we have to apply the multi fluid approach to describe the system equation of motion \citep{RA,Ma,Mag}. Separation of QGP constituents in term of their charges due to strong magnetic field, also has been proven experimentally \citep{s,wang}.   

For describing the magnetohydrodynamics, we can write a system of equations which are governed by the conservation law associated to baryon density, the energy-momentum equations of motion and evolution of the electromagnetic field through the  Maxwell's equations, as following \citep{Apr, Ap,rpm 1,RA}:
 \begin{equation}\label{1}
 \frac{\partial\rho_{Bi}}{\partial t}+{\bf {\nabla}} . (\rho_{Bi} {{\bf{v}}_i})=0
 \end{equation}
\begin{equation}\label{2}
\rho_{mi}\left(\frac{\partial }{\partial t}+{{\bf{v}}_i} . {\bf{\nabla}}\right){{\bf{v}}_i}=  -{\bf{\nabla}} p_i +\rho_{Ci}(
{\bf{E}}+{{\bf{v}}_i}\times {\bf{B}})
\end{equation}
\begin{equation}\label{5}
\frac{\partial {\bf{B}}}{\partial t}=-{\bf{\nabla}}\times{\bf{E}}
\end{equation}
\begin{equation}\label{6}
{\bf{\nabla}}\times{\bf{B}} -\varepsilon\mu\frac{\partial {\bf{E}}}{\partial t}=\mu\sum\limits_{i}  \rho_{Ci}{\bf{v}}_i 
\end{equation}
\begin{equation}\label{7}
{\bf{\nabla}}.{\bf{B}}=0
\end{equation}
\begin{equation}\label{8}
{\bf{\nabla}}.{\bf{E}}=\frac{1}{\varepsilon}\sum \limits_{i} \rho_{Ci}
\end{equation}
Equations (\ref{1}) and (\ref{2}) are the baryon density continuity equation and non relativistic equation of motion, where $ \rho_{mi} $, $ \rho_{B_i} $, $ \rho_{C_i} $ and ${ \bf{v}}_i $ denote respectively, the mass density, baryon density, charge density and  the velocity of each quark with $i= u , d$. Also, the equations (\ref{5})-(\ref{8}) are the Maxwell's equations with the magnetic field $ {\bf{B}} $ and the electric vector field $ {\bf{E}} $, while $\varepsilon$ and $\mu $ are effective dielectric constant and magnetic permeability respectively. If we assume that the net charge is negligible ($\rho_{Cu}=\rho_{Cd}=N$) \citep{Apr,Ap}, by considering the relation between charge density and baryon density of quarks, we obtain $\rho_{B} =\frac{3}{2} N $. Also, by ignoring the displacement current\citep{Ap}, the equation (\ref{6}) becomes:
\begin{equation}\label{6a}
{\bf{\nabla}}\times{\bf{B}}=\mu \frac{2}{3}\rho_{B}({\bf{v}}_u - {\bf{v}}_d)
\end{equation}
Substituting above expression into (\ref{2}), obtaining the electric field in terms of velocity and magnetic field and assuming the fluid velocity as $({\bf{v}} \approx {\bf{v_d}}) $ since the $ d $ quarks carry almost all of the momentum, we obtain:
\begin{eqnarray}\label{11}
\rho_m\frac{d{\bf{v}} }{d t}=&&  - {\bf{\nabla}} p+ \frac{1}{\mu } ( {\bf{\nabla}}\times {\bf{B}})\times {\bf{B}} \nonumber\\
&&- \frac{\rho_{mu}}{\frac{2}{3}\mu\rho_B} \left[ \left\lbrace \left( {\bf{\nabla}}\times{\bf{B}}\right). {\bf{\nabla}}\right\rbrace {{\bf{v}}} + \frac{d}{dt}\left( {\bf{\nabla}}\times{\bf{B}}\right)\right] \nonumber\\ 
&& -\frac{\rho_{mu}}{(\frac{2}{3}\mu\rho_B)^2}\left\lbrace \left({\bf{\nabla}}\times{\bf{B}}\right). {\bf{\nabla}}\right\rbrace \left({\bf{\nabla}}\times{\bf{B}}\right)
\end{eqnarray}
and
\begin{equation}\label{12}
\frac{\partial {\bf{B}}}{\partial t}= {\bf{\nabla}}\times ({{\bf{v}}}\times {\bf{B}}) +\frac{\rho_{md}}{\frac{2}{3}\rho_B}\left( {\bf{\nabla}}\times \frac{d{{\bf{v}}}}{dt}\right) 
\end{equation}
where $ d/dt $ denotes $ \partial/ \partial t +({\bf{v}}.{\bf{\nabla}}) $.
In this step, we have derived a complete set of equations which describe time evolution of the cold QGP system. Unfortunately above equations are highly nonlinear.  We have not allowed to ignore dispersion, dissipation and nonlinear terms as they play essential roles in the dynamics of the system. Luckily, it is possible to investigate the behaviour of the system by considering all above terms using the reductive perturbation method (RPM) for small amplitude excitations \citep{p1, Non-linear,p2}. 

\section*{3. The QGP equation of state}
The equation set (\ref{1})-(\ref{8}) can not be evaluated only if we add another equation, describing the fluid pressure ($p$) respect to other variables of the system, which calls equation of state (EOS). Several equation of ststes have been proposed for QGP through different approaches. The MIT bag model \cite{bag}, strongly interacting QGP model \citep{si}, Cornell potential model \citep{cm} are some of the most famous presented EOS for QGP. It may be noted that there is not any model which is completely accepted by physics community. Most of studies have been done using the MIT bag model as EOS. In order to find comparable results with outcomes of other investigations, we have used this model, however similar procedure can be applied for other forms of EOS.\

 The fundamental idea of the MIT bag model describes the QGP as an ideal gas of non interacting quarks and gluons. Quarks move freely inside the bag to the first order of approximation while their interaction with gluons are not taken into account \citep{rpm 1, RA}. The confinement interpreted through the bag constant $B_{bag} $ which is the needed energy to create a bag in the QCD vacuum. According to the boundary condition due to confinement in the MIT bag model, quarks find very small mass inside the bag. Mass of $ u$ and $d$ quarks in dense quark matter are estimated about $ 5 MeV$  using the MIT bag model \citep{bu} while their mass are infinity at the boundary and/or outside this environment \citep{i22}. Therefore the mass of light quarks are negligible in dense quark matter and quark gluon plasmas. On the other hand the quark effective mass also reduces in strong magnetic fields \citep{me}.\

The baryon density is given by \citep{Th} : 
\begin{equation}\label{B1}
\rho_{B}=\frac{1}{3}\frac{\gamma_{Q}}{(2\pi)^{3}}\int d^{3}k [n_{\vec{k}} - \bar{n}_{\vec{k}}]
\end{equation}
where $n_{\vec{k}}$ and $\bar{n}_{\vec{k}}$ are quark and anti quark distribution functions which are given by the Fermi-Dirac distribution function in dense quark matter:
\begin{equation}\label{N}
n_{\vec{k}}\equiv n_{\vec{k}}(T)= \frac{1}{1+e^{( k -\frac{1}{3}\mu_{ch})/T}}
\end{equation}
and
\begin{equation}\label{Nbar}
\bar{n}_{\vec{k}}\equiv \bar{n}_{\vec{k}}(T)= \frac{1}{1+e^{(k+\frac{1}{3}\mu_{ch})/T}}
\end{equation}
in which $\mu_{ch}$ is the baryon chemical potential. The energy density and the pressure are given by \citep{i23}: 
\begin{equation}\label{E}
\varepsilon= B_{bag}+\frac{\gamma_{G}}{(2\pi)^{3}}\int d^{3}k \:k \:(e^{k /T}-1)^{-1}+\frac{\gamma_{Q}}{(2\pi)^{3}}\int d^{3}k \:k \:[n_{\vec{k}} + \bar{n}_{\vec{k}}]
\end{equation}
\begin{equation}\label{P}
p= -B_{bag}+\frac{1}{3}\left( \frac{\gamma_{G}}{(2\pi)^{3}}\int d^{3}k \:k \:(e^{k /T}-1)^{-1}+\frac{\gamma_{Q}}{(2\pi)^{3}}\int d^{3}k \:k \:[n_{\vec{k}} + \bar{n}_{\vec{k}}]\right). 
\end{equation}
Above quantities are calculated by taking just two flavours of quarks $  (u,d)$, so that, the degeneracy factors are $\gamma_{G}=16$ for gluons and $\gamma_{Q}=12$ for quarks. By considering the zero temperature, the contribution of gluons in the energy and momentum became zero.  So we have:
\begin{equation}
p=\frac{1}{3}\varepsilon - \frac{4}{3} B_{bag}
\end{equation}
and the speed of sound, $v_{S}$ is given by:
\begin{equation}
v_{S}^2=\frac{\partial p}{\partial\varepsilon}=\frac{1}{3}
\end{equation}
It may be noted that the quarks distribution function at zero temperature is the step function. Such this medium as cold QGP can be found in the core of super dense astrophysical objects, which their temperatures are close to zero \citep{Non-linear}. At zero temperature the expression for the baryon density (\ref{B1}) becomes:
\begin{equation}\label{B2}
\rho_{B}=\frac{2}{3\pi^{2}}k_{F}^{3}
\end{equation}
where $k_{F}$ is the highest occupied momentum level. Using (\ref{B2}) in (\ref{E}) and (\ref{P}) we can write the energy density and pressure as following:
\begin{equation}\label{E1}
\varepsilon(\rho_{B})=\left( \frac{3}{2}\right) ^{7/3}\pi^{2/3} \rho_{B}^{4/3} +B_{bag}
\end{equation}
\begin{equation}\label{P1}
p(\rho_{B})=\frac{1}{3}\left(\frac{3}{2}\right)^{7/3}\pi^{2/3} \rho_{B}^{4/3} -B_{bag}
\end{equation}
From (\ref{P1}) we have:
\begin{equation}\label{Gr}
\vec{\nabla}p= \frac{4}{9}\left(\frac{3}{2}\right)^{7/3}\pi^{2/3} \rho_{B}^{1/3}\vec{\nabla}\rho_{B}
\end{equation}
In the non relativistic limit $\varepsilon+p\cong\rho$ \citep{rpm 1, viscose} and thus:
\begin{equation}\label{Nrho}
\rho=\frac{4}{3}\left(\frac{3}{2}\right)^{7/3}\pi^{2/3} \rho_{B}^{4/3}.
\end{equation}
The above equations, (\ref{Gr}) and (\ref{Nrho}), can be used in magnetohydrodynamic equations as the equation of state.\

Another version of MIT bsg model with negative bag constant has been presented \citep{nb}. This new EOS satisfies needed conditions provided in the lattice QCD simulations. The energy density in this model also is similar to that in the standard bag model, but with a negative value of the bag constant. Therefore one can consider modified MIT bag model in calculations.  \\
In the next section, we present time evolution equations of the system by introducing suitable variables using the RPM formalism.

\section*{4. The reductive perturbation method}
Consider infinitely extended uniforme cold QGP which its physical quantities at equilibrium state are given by\citep{p4}:
\begin{eqnarray}
\rho_B=\rho_{B0}  \qquad {\bf{B}}={B}_0 \hat{e}_x \qquad {{\bf{v}}}=0 \nonumber
\end{eqnarray}
Components of velocity and magnetic field vectors are ${{\bf{v}}}=(v_{x}, v_{y},v_{z})$ and $ {\bf{B}}=(B_x,B_y,B_z) $. Transverse components of velocity ($\tilde{v}  $) and magnetic field ($ \tilde{B} $) can be introduced by following complex quantities:
\begin{eqnarray}\label{13}
\tilde{v} = v_{y}+ iv_{z}  \qquad  \qquad  \tilde{B}= B_y+iB_z
\end{eqnarray}
We take the following variables as the stretched coordinates:
\begin{eqnarray}\label{14}
&&\xi=\epsilon(x-\lambda t)\qquad\qquad \eta=\epsilon^{3/2}y\nonumber\\
&&\zeta=\epsilon^{3/2}z \qquad\qquad\qquad \tau=\epsilon^2 t
\end{eqnarray}
where $ \lambda $ is the wave velocity at linear approximation and $ \epsilon $ is the expansion parameter, i.e. $ \epsilon<1 $.
Here, we can expand the variables in power series of $ \epsilon $ as follows:
\begin{eqnarray}\label{15}
&&\rho_B=\rho_{B0} +\epsilon \rho^{(1)}_B+...\nonumber\\
&&\tilde{B}=\epsilon^{1/2}(\tilde{B}^{(1)}+\epsilon\tilde{B}^{(2)}+...)\nonumber\\
&&\tilde{v} =\epsilon^{1/2}(\tilde{v} ^{(1)}+\epsilon\tilde{v} ^{(2)}+...)\nonumber\\
&&B_x=B_0+\epsilon B_x^{(1)}+...\nonumber\\
&&v_{x}=\epsilon v_{x}^{(1)}+\epsilon^2 v_{x}^{(2)}+...\nonumber\\
&&p=p_{0} +\epsilon p^{(1)}+...
\end{eqnarray}

We now use the stretched coordinates (\ref{14}) and expansions (\ref{15}) in (\ref{1}),
(\ref{11}) and (\ref{12}). At the lowest order of $ \epsilon $, following results are obtained:
\begin{eqnarray}\label{16}
-A\rho_{B0}^{4/3}\lambda \frac{\partial v_{y}^{(1)}}{\partial \xi} = \frac{1}{\mu}B_0 \frac{\partial B_{y}^{(1)}}{\partial\xi }&&\qquad -A\rho_{B0}^{4/3}\lambda \frac{\partial v_{z}^{(1)}}{\partial\xi} = \frac{1}{\mu}B_0 \frac{\partial B_{z}^{(1)}}{\partial\xi}\nonumber\\
-\lambda  \frac{\partial B_{y}^{(1)}}{\partial\xi}= B_0 \frac{\partial v_{y}^{(1)}}{\partial\xi}&&\qquad -\lambda  \frac{\partial B_{z}^{(1)}}{\partial\xi}= B_0 \frac{\partial v_{z}^{(1)}}{\partial\xi}
\end{eqnarray}
Above equations can be written as:
\begin{equation}\label{17}
A_\lambda \frac{\partial}{\partial\xi} \left(\begin{array}{cc} \tilde{v} ^{(1)}\\ \tilde{B}^{(1)}\end{array} \right)= \left(\begin{array}{cc} 0\\ 0\end{array} \right)
\end{equation}
where 
\begin{equation}\label{18}
A_\lambda=\left( \begin{array}{cc}
-\lambda &-\frac{v_A^2}{B_0}\\
-B_0&-\lambda
\end{array}\right) 
\end{equation}
and $ v_A $ is the Alfven velocity which can be derived as:
\begin{equation}\label{19}
v_A^2=\frac{B_0^2}{\frac{4}{3}\left(\frac{3}{2}\right)^{7/3}\pi^{2/3} \mu  \rho_{B0}^{4/3}}
\end{equation}
where $ \frac{4}{3}\left(\frac{3}{2}\right)^{7/3}\pi^{2/3} \rho_{B0}^{4/3} $ is  the mass density according to the equation (\ref{Nrho}). Also from the equation (\ref{17}) we have:
\begin{eqnarray}\label{20}
\lambda= v_A\qquad \qquad,\qquad\qquad \tilde{v} ^{(1)}= - \frac{v_A}{B_0} \tilde{B}^{(1)}
\end{eqnarray}
At the order of $ \epsilon^2 $, from the equations (\ref{1}), (\ref{11}) and (\ref{12}) we have:
\begin{eqnarray}\label{21}
-\frac{\lambda}{\rho_{B0}} \frac{\rho_{B}^{(1)}}{\partial \xi} + \frac{\partial v_{x}^{(1)}}{\partial \xi} + \frac{\partial v_{y}^{(1)}}{\partial \eta}+\frac{\partial v_{z}^{(1)}}{\partial \zeta}=0
\end{eqnarray}
\begin{equation}\label{22}
\lambda A \rho_{B0}^{4/3} \frac{\partial v_{x}^{(1)}}{\partial \xi}=\frac{1}{3} A \rho_{B0}^{1/3} \frac{\rho_{B}^{(1)}}{\partial \xi} + \frac{1}{\mu_0}\left[ B_{y}^{(1)}\frac{\partial B_{y}^{(1)}}{\partial \xi}+B_{z}^{(1)}\frac{\partial B_{z}^{(1)}}{\partial \xi} \right]
\end{equation}
\begin{equation}\label{23}
\lambda  \frac{\partial B_{x}^{(1)}}{\partial \xi} = B_0 \frac{\partial v_{y}^{(1)}}{\partial \eta}+ B_0 \frac{\partial v_{z}^{(1)}}{\partial \zeta}.
\end{equation}
From equations (\ref{21}) and (\ref{23}) we obtain:
\begin{equation}\label{24}
-\lambda\frac{\partial \rho_{B}^{(1)}}{\partial \xi} + \rho_{B0} \frac{\partial v_{x}^{(1)}}{\partial \xi}+ \rho_{B0} \frac{\lambda}{B_0} \frac{\partial B_{x}^{(1)}}{\partial \xi}=0
\end{equation}
which result in:
\begin{equation}\label{25}
\frac{\rho_{B}^{(1)}}{\rho_{B0}}=\frac{v_{x}^{(1)}}{v_A}+\frac{B_{x}^{(1)}}{B_0}
\end{equation}
and
\begin{equation}\label{26}
 \frac{\partial B_{x}^{(1)}}{\partial \xi}+{\bf{\nabla_\perp}}. {\bf{B}} ^{(1)}_\perp=0
\end{equation}
where
\begin{equation}\label{27}
{\bf{\nabla_\perp}}=(\frac{\partial}{\partial\eta}, \frac{\partial}{\partial\zeta})\qquad \qquad,\qquad\qquad {\bf{B}} _\perp=(B_y , B_z).
\end{equation}
Also, from (\ref{22}) and (\ref{25}) we arrive at:
\begin{equation}\label{28}
v_{x}^{(1)}=\frac{v_A}{v_A^2 - v_S^2}\left[ v_S^2 \frac{B_{x}^{(1)}}{B_0} +\frac{v_A^2}{2} \frac{\mid \tilde{B}^{(1)} \mid^2}{B_0^2}\right] 
\end{equation}
where $ v_S $ is the sound velocity which is appeared in the equation of state.

From the terms of the order $ \epsilon^{5/2} $, following equations are derived:
\begin{eqnarray}\label{29}
&&A\rho_{B0}^{4/3}\left( \frac{\partial v_{y}^{(1)} }{\partial \tau} - v_A \frac{\partial v_{y}^{(2)} }{\partial \xi} +v_{x}^{(1)} \frac{\partial v_{y}^{(1)} }{\partial \xi} \right) - \frac{4}{3} A v_A\rho_{B0}^{1/3} \frac{\partial v_{y}^{(1)} }{\partial \xi}\nonumber\\
&&+ v_{y}^{(1)} \frac{\partial v_{y}^{(1)} }{\partial \eta}+v_{z}^{(1)} \frac{\partial v_{y}^{(1)} }{\partial \zeta}=-\frac{1}{3} A\rho_{B0}^{1/3} \frac{\partial \rho_B^{(1)}}{\partial \eta} - \frac{A \rho_{B0}^{1/3}}{2\mu} v_A \frac{\partial ^2 B_z^{(1)}}{\partial \xi^2}\nonumber\\
&&+\frac{1}{\mu} \left[ B_0 \frac{\partial B_y^{(2)}}{\partial \xi} + B_x^{(1)} \frac{\partial B_y^{(1)}}{\partial \xi} +B_z^{(1)} \frac{\partial B_y^{(1)}}{\partial \zeta} - B_0 \frac{\partial B_x^{(1)}}{\partial \eta} - B_z^{(1)} \frac{\partial B_z^{(1)}}{\partial \eta}\right] \nonumber\\
\end{eqnarray}

\begin{eqnarray}\label{30}
&&A\rho_{B0}^{4/3}\left( \frac{\partial v_{z}^{(1)} }{\partial \tau} - v_A \frac{\partial v_{z}^{(2)} }{\partial \xi} +v_{x}^{(1)} \frac{\partial v_{z}^{(1)} }{\partial \xi} \right) - \frac{4}{3} A v_A\rho_{B0}^{1/3} \frac{\partial v_{z}^{(1)} }{\partial \xi}\nonumber\\
&&+ v_{uy}^{(1)} \frac{\partial v_{uz}^{(1)} }{\partial \eta}+v_{uz}^{(1)} \frac{\partial v_{uz}^{(1)} }{\partial \zeta}=-\frac{1}{3} A\rho_{B0}^{1/3} \frac{\partial \rho_B^{(1)}}{\partial \eta} + \frac{A \rho_{B0}^{1/3}}{2\mu} v_A \frac{\partial ^2 B_y^{(1)}}{\partial \xi^2}\nonumber\\
&&+\frac{1}{\mu} \left[ B_0 \frac{\partial B_z^{(2)}}{\partial \xi} + B_x^{(1)} \frac{\partial B_z^{(1)}}{\partial \xi} +B_y^{(1)} \frac{\partial B_z^{(1)}}{\partial \eta} - B_0 \frac{\partial B_x^{(1)}}{\partial \zeta} - B_y^{(1)} \frac{\partial B_y^{(1)}}{\partial \zeta}\right] \nonumber\\
\end{eqnarray}

\begin{eqnarray}\label{31}
\frac{\partial B_y^{(1)}}{\partial \tau} - v_A \frac{\partial B_y^{(2)}}{\partial \xi} = &&\frac{\partial }{\partial \zeta} \left( v_{y}^{(1)} B_z^{(1)} - v_{z}^{(1)} B_y^{(1)} \right) +B_0 \frac{\partial v_{y}^{(2)} }{\partial \xi}\nonumber\\
 &&\frac{\partial }{\partial \xi} \left( v_{x}^{(1)} B_y^{(1)} - v_{y}^{(1)} B_x^{(1)} \right) + A \rho_{B0}^{1/3} v_A \frac{\partial ^2 v_{z}^{(1)}}{\partial \xi^2}\nonumber\\
\end{eqnarray}

\begin{eqnarray}\label{32}
\frac{\partial B_z^{(1)}}{\partial \tau} - v_A \frac{\partial B_z^{(2)}}{\partial \xi} = &&\frac{\partial }{\partial \xi} \left( v_{z}^{(1)} B_x^{(1)} - v_{x}^{(1)} B_z^{(1)} \right) +B_0 \frac{\partial v_{z}^{(2)} }{\partial \xi}\nonumber\\
 &&-\frac{\partial }{\partial \eta} \left( v_{y}^{(1)} B_z^{(1)} - v_{z}^{(1)} B_y^{(1)} \right) +A \rho_{B0}^{1/3} v_A \frac{\partial ^2 v_{y}^{(1)}}{\partial \xi^2}.\nonumber\\
\end{eqnarray}

Above equations can be written as following, using (\ref{20}), (\ref{25}) and (\ref{26}):
\begin{eqnarray}\label{33}
&&\frac{\partial \tilde{v}^{(1)}}{\partial \tau} - v_A \frac{\partial \tilde{v}^{(2)}}{\partial \xi} - \frac{v_A^2}{B_0} \frac{\partial \tilde{B}^{(2)}}{\partial \xi} + \frac{v_S^2 v_A^4}{B_0(v_A^2 -v_S^2)}\left( \frac{B_x^{(1)}}{B_0} + \frac{\mid \tilde{B}^{(1)} \mid^2}{2B_0^2} \right) \frac{\partial \tilde{B}^{(1)}}{\partial \xi}= \nonumber\\
&& -{\bf{\tilde{\nabla}}}\left[\frac{v_A^4}{v_A^2 -v_S^2} \left(  \frac{B_x^{(1)}}{B_0} + \frac{\mid \tilde{B}^{(1)} \mid^2}{2B_0^2}\right)  \right] + i \frac{v_A}{2 \mu \rho_{B0}} \frac{\partial ^2\tilde{B}^{(1)}}{\partial \xi^2}
\end{eqnarray}
\begin{eqnarray}\label{34}
\frac{\partial \tilde{B}^{(1)}}{\partial \tau} - B_0 \frac{\partial \tilde{v}^{(2)}}{\partial \xi} - v_A\frac{\partial \tilde{B}^{(2)}}{\partial \xi} = \frac{\partial }{\partial \xi} \left( \tilde{v}^{(1)} B_x^{(1)} - \tilde{B}^{(1)} v_{ux}^{(1)} \right) - i A \rho_{B0}^{1/3} v_A \frac{\partial ^2 \tilde{v}^{(1)}}{\partial \xi^2}\nonumber\\
\end{eqnarray}
$\tilde{v}^{(2)}, \tilde{B}^{(2)}, \tilde{v}^{(1)} $ and $  v_{x}^{(1)} $ can be eliminated from equations (\ref{33}) and (\ref{34}) with using equations (\ref{20}) and (\ref{28}). Thus we can derive the master equation containing $ \tilde{B}^{(1)} $, $ B_x^{(1)}$ and their derivatives as following:
\begin{eqnarray}\label{35}
\frac{\partial \tilde{B}^{(1)}}{\partial \tau} - B_0 {\bf{\tilde{\nabla}}} \Upsilon + \frac{\partial }{\partial \xi} \left(\Upsilon \tilde{B}^{(1)} \right) - v_S^2 \Upsilon  \frac{\partial \tilde{B}^{(1)}}{\partial \xi} + i C_1 \frac{\partial ^2\tilde{B}^{(1)}}{\partial \xi^2}=0
\end{eqnarray}
where 
\begin{equation}\label{36}
\Upsilon = \frac{1}{2} \frac{v_A^3}{v_A^2 -v_S^2} \left(  \frac{B_x^{(1)}}{B_0} + \frac{\mid \tilde{B}^{(1)} \mid^2}{2B_0^2}\right) 
\end{equation}
and
\begin{equation}\label{37}
C_1 = -\frac{B_0}{4\mu \rho_{B0}}.
\end{equation}
Relations (\ref{26}) and (\ref{35}) are the appropriate set of equations governing the evolution of the transverse ($ \tilde{B}^{(1)} $)
and longitudinal ($ B_x^{(1)}) $ magnetic field perturbations. If the spatial variation in the transverse direction is negligible \citep{p4}, the equations (\ref{26}) and (\ref{35}) reduce to the following single equation:
\begin{eqnarray}\label{38}
\frac{\partial \tilde{B}^{(1)}}{\partial \tau} +\frac{1}{2}C_2 \frac{\partial }{\partial \xi} \left(\mid \tilde{B}^{(1)} \mid^2 \tilde{B}^{(1)} \right) + i C_1 \frac{\partial ^2\tilde{B}^{(1)}}{\partial \xi^2}- v_S^2 C_2 \mid \tilde{B}^{(1)} \mid^2\frac{\partial \tilde{B}^{(1)}}{\partial \xi} =0\nonumber\\
\end{eqnarray}
in which:
\begin{eqnarray}\label{39}
C_2=\frac{1}{2} \frac{v_A^3}{v_A^2 -v_S^2}
\end{eqnarray}
The above wave equation in the Cartesian coordinates $ (x,t) $ becomes:
\begin{eqnarray}\label{38a}
&&\frac{\partial \hat{\tilde{B}} ^{(1)} }{\partial t} + v_A \frac{\partial \hat{\tilde{B}} ^{(1)}}{\partial x}+\frac{1}{2}C_2 \frac{\partial }{\partial x} \left(\mid \hat{\tilde{B}}^{(1)} \mid^2 \hat{\tilde{B}}^{(1)} \right)\nonumber\\
&&+ i C_1 \frac{\partial ^2 \hat{\tilde{B}}^{(1)}}{\partial x^2}- v_S^2 C_2 \mid \hat{\tilde{B}}^{(1)}\mid^2\frac{\partial \hat{\tilde{B}}^{(1)}}{\partial x} =0
\end{eqnarray}
while $ \hat{\tilde{B}} ^{(1)} \equiv \sigma ^{1/2} \tilde{B}^{(1)} $.
The equation (\ref{38a}) describes the evolution of transverse magnetic field perturbation. The evolution of baryon density perturbation also can be derived by applying the (\ref{25}) and (\ref{28}) in (\ref{38a}). 
The above equation can be compared with the following version of the complex Ginzburg-Landau equation:
\begin{eqnarray}\label{40}
\frac{\partial u}{\partial t} = iP \frac{\partial ^2 u}{\partial x ^2} +i\gamma u +iQ_1 \mid u \mid ^4 u +Q_2 \mid u\mid ^2 \frac{\partial u}{\partial x } +Q_3 u^2 \frac{\partial \tilde{u}}{\partial x }
\end{eqnarray}
where all coefficients are real and $ u $ is a complex function of space and time $(x,t) $. Equation (\ref{40}) is called the derivative nonlinear Schrodinger (DNLS) equation with an additional potential, or the cubic-quintic Ginzburg-Landau equation \citep{p5}. Two terms $ \mid u\mid ^2 \partial u /\partial x  $ and $ u^2 \partial \tilde{u}/ \partial x  $, are nonlinear dispersion terms. These two terms can significantly reduce the speed of the wave pulse and deform the profile of the wave into non symmetric shapes \citep{p6}. The solution of the DNLS equation has been derived and discussed by several authors \citep{p7,p8,p10}. The equation (\ref{38a}) also cantaining two additional terms: $ \mid \hat{\tilde{B}}^{(1)}\mid^2 \partial \hat{\tilde{B}}^{(1)}/\partial x $ and $ \partial \hat{\tilde{B}}^{(1)}/\partial x $, in comparison with the original DNLS equation \citep{p8,p10}. So we call this equation as the modified derivative nonlinear Schrodinger (mDNLS) equation. In the next section, we derive exact solitonic solutions using the plane wave perturbation technique applied on the equation (\ref{38a}) \citep{p5}.

\section*{5. Solitary wave solution for the mDNLS equation }
In order to derive localizae solutions of the (mDNLS) equation, firstly we express $ \hat{\tilde{B}}^{(1)} (x,t) $ in polar coordinate as following:
\begin{equation}\label{41}
\hat{\tilde{B}}^{(1)} (x,t)= a(x,t) e^{i\theta (x,t) }
\end{equation}
where $a(x,t)$ and $\theta(x,t) $ are real functions. Substituting (\ref{41}) into (\ref{38a}) and separating real and imaginary parts, we have:
\begin{eqnarray}\label{42}
a\frac{\partial \theta}{\partial t}+C_1\left( \frac{\partial ^2 a}{\partial x ^2} - a (\frac{\partial \theta }{\partial x })^2\right)  +v_A a \frac{\partial \theta}{\partial x }+\frac{1}{6}C_2 a^3 \frac{\partial \theta}{\partial x }=0\nonumber\\ 
-\frac{\partial a}{\partial t} +C_1\left( 2 \frac{\partial \theta}{\partial x } \frac{\partial a}{\partial x } + \frac{\partial ^2 \theta}{\partial x ^2}\right) - v_A \frac{\partial a}{\partial x }- \frac{7}{6} C_2 a^2 \frac{\partial a}{\partial x }=0
\end{eqnarray}
The Stokes solution of the above equations is obtained as:
\begin{equation}\label{43}
\hat{\tilde{B}}^{(1)} (x,t)= a_0 e^{i\left[ l_0 x -(\frac{1}{6}C_2 l_0 a_0^2-C _1 l_0^2 - v_A)t \right] }.
\end{equation}
that we use following expressions for parameters $ a $ and $ \theta $ in (\ref{41}):
\begin{equation}\label{44}
a= a_0     \qquad   , \qquad  \theta= l_0 x - q_0 t
\end{equation}
where $ q_0 = \frac{1}{6}(C_2 l_0 a_0^2)-C _1 l_0^2 +v_A l_0 $ in which $ a_0 $ and $ l_0 $ are real constants.
In the next step, we find a solution for the system of equations (\ref{42}) by pretreating the nontrivial solutions (\ref{43}) as follows:
\begin{eqnarray}\label{45}
a(x , t) = a_0 + \alpha (x - v t =\vartheta)\nonumber\\
\theta (x , t)= \Psi (\vartheta) - (q_0 - l_0 v) t
\end{eqnarray}
where $ v $ is a constant parameter. Inserting  (\ref{45}) into (\ref{42}) we get:
\begin{eqnarray}\label{46}
&&- v a \Psi ^ \prime +C_1\left( \alpha ^{\prime \prime} - a (\Psi ^ \prime)^2 \right) + \frac{1}{6} C_2 a^3 \Psi ^ \prime+ v_A a \Psi ^ \prime + (l_0 v - q_0) a=0\nonumber\\
&&(v - v_A ) \alpha ^ \prime + C_1\left( 2 \alpha ^ \prime \Psi ^ \prime + a \Psi ^ {\prime\prime} \right) -\frac{7}{6} a^2 \alpha ^ \prime =0
\end{eqnarray}
Multiplying the second equation of the set (\ref{46}) by $ a $ and integrating the resulting equation yields:
\begin{equation}\label{47}
\Psi ^ \prime = \frac{K_1}{C_1 a^2} + \frac{7 C_2}{24 C_1} a^2 - \frac{v - v_A }{2 C_1}
\end{equation}
where $ K_1 $ is a constant of integration. Substituting  the above equation into the first equation (\ref{46}), we obtain:
\begin{eqnarray}
a^{\prime \prime} =&& \frac{1}{4 C_1^2}\left[ \frac{5 C_2}{3} K _1 - (v - v_A)  ^2 - 4 C_1 (l_0 v - q_0) \right] a \nonumber\\ 
&&+\frac{ C_2}{12 C_1^2} (v - v_A) a^3 + \frac{K_1^2}{C_1^2 a^3} + \frac{7 C_2}{12(4C_1)^2} a^5
\end{eqnarray}
the first integral of above equation gives:
\begin{eqnarray}\label{48}
{a^\prime}^ 2 =&& \frac{1}{4 C_1^2} \left[  \frac{5 C_2}{3} K _1 - (v - v_A) ^2 - 4 C_1 (l_0 v - q_0) \right] a^2 \nonumber\\
&&+ \frac{ C_2}{24 C_1^2} (v - v_A) a^4 +\frac{K_2}{4}+  \frac{7 C_2}{(48 C_1)^2} a^6 - \frac{K_1^2}{C_1^2 a^2}
\end{eqnarray}
where $ K_2 $  is a constant of integration. By setting $ {a}^ 2 = \chi $ we obtain an elliptic ordinary differential equation as following : 
\begin{eqnarray}\label{49}
{\chi ^\prime}^2 = \frac{-4 K_1 ^2}{C_1^2} +K_2 \chi +D \chi ^2 + E \chi ^3 + F \chi ^4
\end{eqnarray}
where
\begin{eqnarray}\label{50}
D = \frac{ \frac{5 C_2}{3} K _1 - (v - v_A) ^2 - 4 C_1 (l_0 v - q_0)}{C_1^2}\nonumber\\
E = \frac{ C_2}{6 C_1^2} (v - v_A) \qquad, \qquad F= \frac{28 C_2}{(48 C_1)^2}
\end{eqnarray}
Now, we can examine the behaviour of solitary solutions for this equation by considering fixed values for free parameters. When  $ K_1 =  K_2 =0 $, the equation (\ref{49}) becomes:
\begin{eqnarray}\label{51}
\frac{\chi ^{\prime^2}}{\chi^2} = D+E \chi + F \chi ^2 
\end{eqnarray}
In this case, if coefficients $ D $ ,$ E $ and $ F $  satisfy the following conditions:
\begin{equation}\label{51a}
D > 0  \qquad and  \qquad E^2 - 4DF > 0
\end{equation}
the equation (\ref{49}) embraces the following solitonic solution:
\begin{eqnarray}\label{52}
\chi(\vartheta) = \chi^{\pm}(\vartheta) =  \frac{2D}{-D \pm \sqrt{E^2 - 4DF} \cosh \left( \sqrt{D} \vartheta\right) }
\end{eqnarray}
Since $ {a}^ 2 = \chi $ and $ x - v t =\vartheta $, from (\ref{41}) we can write :
\begin{eqnarray}\label{53}
\mid \hat{\tilde{B}}^{(1)} (x,t)\mid ^ 2=  \frac{2D}{-D \pm \sqrt{E^2 - 4DF} \cosh \left( \sqrt{D} (x - vt)\right) }
\end{eqnarray}
This equation clearly shows that the transverse magnetic field perturbation (and consequently baryon density perturbation) propagates as stable solitary waves in cold QGP environment. The constraint $ D > 0 $ leads to the $  (v - v_A) ^2 + 4 C_1 (l_0 v - q_0) <0 $, so that we can determine the propagation velocity of solitary wave perturbation. It may be noted that selecting the free parameters $ a_0  ,  l_0 $ causes a convenient range for speed of localized wave that is greater than the Alfven velocity.

Now we are able to calculate the electric field in the QGP. From the Maxwell equation (\ref{5}) we find that the direction of the electric field perturbation is along the initial magnetic field ( $x$ direction). Using (\ref{53}) and (\ref{45}) in (\ref{41}) and considering the Maxwell equation (\ref{5}), the electric field perturbation can be obtained as following:
\begin{eqnarray}\label{54}
&&\mid \hat{E}_x^{(1)} (x,y,z,t)\mid ^2= \left[(z-y)\frac{0.7 D v \sqrt{E^2 - 4DF} \sinh \left( \sqrt{D} (x - vt)\right)}{\left\lbrace  -D \pm \sqrt{E^2 - 4DF} \cosh( \sqrt{D} (x - vt))\right\rbrace  ^{3/2}}\right] ^2 + \nonumber\\
&& \frac{ (y-z)^2 2D}{-D \pm \sqrt{E^2 - 4DF} \cosh \left( \sqrt{D} (x - vt)\right) } \left(l_0 v - q_0 - \frac{7 C_2v}{24 C_1} a^2 - \frac{v(v - v_A) }{2 C_1}\right) ^2 + \nonumber\\
&& 2zy\frac{1.4 D^2 v \sqrt{E^2 - 4DF} \sinh \left( \sqrt{D} (x - vt)\right)}{\left\lbrace  -D \pm \sqrt{E^2 - 4DF} \cosh( \sqrt{D} (x - vt))\right\rbrace  ^{2}} \left(l_0 v - q_0 - \frac{7 C_2v}{24 C_1} a^2 - \frac{v(v - v_A) }{2 C_1}\right)\nonumber\\
\end{eqnarray}

It is an important result. existence of this electric field causes a separation between plasma components according to their charges along the magnetic field. This means that we expect to have an electric dipole moment in the core of dense and compact astrophysical objects.

\section*{6. Discussion  }
 Now we are able to study all the features of propagating waves in the cold QGP. The derived localized solution helps us to describe effects of different parameters of the medium on the characteristics of solitary waves propagating in the medium, based on the latest available information about the quark matter and QGP. Equations (\ref{53}) clearly indicate that the wave phase speed, amplitude and  the width of magnetic field perturbations are functions of $\rho_{0} $ (or $\rho_{B0}$ ), $ |B_0| $ and $\mu$, which are the most important characteristics of fluid environment. 

Based on the latest available information about the cold QGP environments, such as neutron stars or quark stars,  we have chosen $\rho_{0}\sim3 \times 10^{17} Kg/m^3$ and $|B_0|=10^{10} T $ \citep{RA, p13}. Also we take the magnetic permeability equal to $\mu _0 = 4\pi \times 10^{-7} H/m $ \citep{p14}. Considering above mentioned values, we can  obtain  $v_A $  (from \ref{19}) , $C_1 $ ( by \ref{37}) , $C_2 $ (using \ref{39})  and consequently other parameters.

At first, we pay attention to the propagation velocity of solitary wave perturbation. If the free parameters of the solution are taken as: $ a_0 \sim 4 $ and $ l_0 \sim 1 $, the constraint $ (v - v_A) ^2 + 4 C_1 (l_0 v - q_0) <0 $ determines a convenient range for the solitary wave speed, which is definitely greater than the Alfven velocity. After selecting a specific value for the velocity within the range of allowed values, the width and amplitude of the localized wave are calculated. Therefore all properties of solitary wave are functions of  background mass density $ \rho_0$ (or baryon density $\rho_{B0})$ and  magnetic field $ |B_0| $ as characteristic parameters of the QGP environment. Derived relations show that the Alfven velocity $ v_A $ increases as the value of $ |B_0| $ increases. In the table 1, we write down the acceptable rang of solitary wave speed by considering different value of background magnetic field $|B_0| $ which have been calculated using the constraint $ D > 0 $. In our solutions, both Alfven wave speed ($v_A$) and perturbation wave velocity are smaller than the sound wave speed: $v_S = \frac{1}{\sqrt{3}} c $. 
\begin{table}
\begin{center}
\begin{tabular}{ c   c   c }
\hline \hline
$ B_0 (T) $ &    Alfven wave velocity (m/s) &  The range of solitary wave velocity (m/s) \\ \hline
$ 1.0\times10^9 $&	 1629.088 & 1767.906 $ - $ 1767.919\\
$ 4.0\times10^{10} $&	66792.609 & 72472.161  $ - $ 72469.714\\
$ 6.0\times10^{10} $&	99374.369 & 107815.369 $ - $ 107869.879 \\
$ 4.0\times10^{11} $&$ 6.532\times10^5 $	& $ 7.076 \times 10^5 - 7.104 \times 10^5$\\
$ 6.0\times10^{11} $&	$ 9.791 \times10^5 $& $ 1.052 \times 10^6 - 1.061 \times 10^6$\\
$ 1.0\times10^{12} $&	$ 1.63 \times10^6 $ & $1.763 \times 10^6 - 1.776 \times 10^6$\\\hline
\end{tabular}
\end{center}
\caption{Alfven wave velocity and the range of solitary wave speed by considering different value of background magnetic field $|B_0| $ at$\rho_0 = 3.0\times 10^{17} (kg/m^3)$. }
\end{table}

Figure 1 demonstrates the time evolution of transverse magnetic field perturbation which propagates without any distortion in its initial direction.  It is clear that such waves are able to reach the border of the medium and create measurable effects at the boundaries. The velocity of solitary wave has been taken as $ v=1.76 \times 10^4 (m/s)$ by considering the Alfven velocity $ v_A = 1.62 \times 10^4 (m/s) $.  Applying the above mentioned values on (\ref{53}) gives the width and amplitude as $ \triangle =(9.51\times10^{3}) (1/m)$ and  $ A= 0.32 B_0^2 $. Also the time evolution of the baryon density perturbation is plotted in figure 2, through applying (\ref{25}) and (\ref{28}) in (\ref{38a}). As this figure presents, the baryon density solitary wave perturbation propagates with a negative amplitude. But its width and velocity is simillar to the transverse magnetic field localized wave .
\begin{figure}[htp]\label{fig1}
\centerline{\begin{tabular}{cc}
\includegraphics[width=10 cm, height= 10 cm]{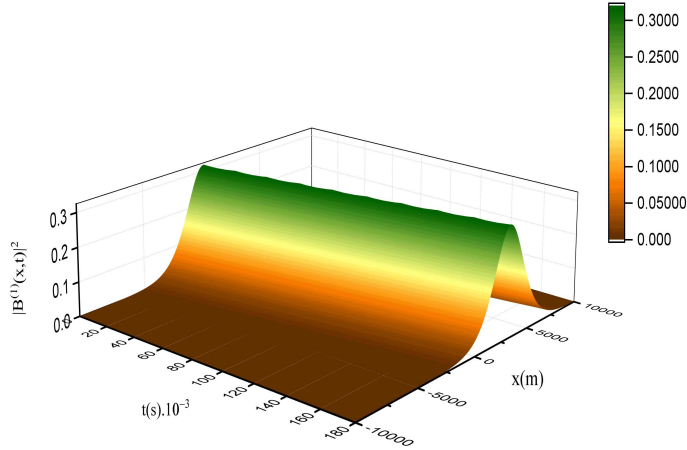}
\end{tabular}}
 \caption{\footnotesize
 Time evolution of the transverse magnetic field perturbation in a cold QGP with $\rho_{0}=3\times 10^{17} kg/m^3$ and $| B_0 |= 10^{10} T  $ . }
\end{figure}
\begin{figure}[htp]\label{fig2}
\centerline{\begin{tabular}{cc}
\includegraphics[width=10 cm, height= 10 cm]{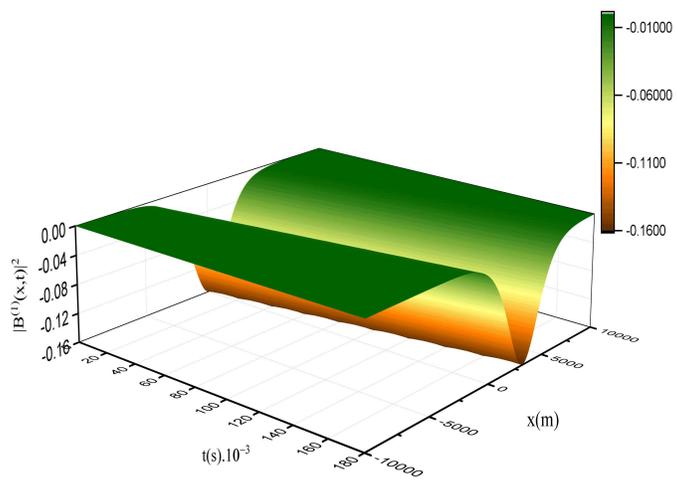}
\end{tabular}}
 \caption{\footnotesize
 Time evolution of the baryon density perturbation in a cold QGP with $\rho_{0}=3\times 10^{17} kg/m^3$ and $| B_0 |= 10^{10} T  $ . }
\end{figure}

As mentioned before, the width and the amplitude of the solitary wave are complicated functions of the background mass density $ \rho_0$ and  magnetic field $ |B_0| $. Figure 3 presents the motion of solitonic profiles in media with the same mass density, but different values of the background magnetic field $| B_0 |$. This figure shows that the width of localized wave (the soliton velocity) decreases (increases), when the strength of magnetic field increases. It also clearly indicates that the amplitude of the solitary wave has not significant changes when the background magnetic field is changed. On the other hand, decreasing the value of $ |B_0| $ causes  an increase in the width of solitary wave and consequently such localized waves can be created in a specific range of magnetic field $| B_0| $.

\begin{figure}[htp]\label{fig3}
\centerline{\begin{tabular}{cc}
\includegraphics[width=10 cm, height=10 cm]{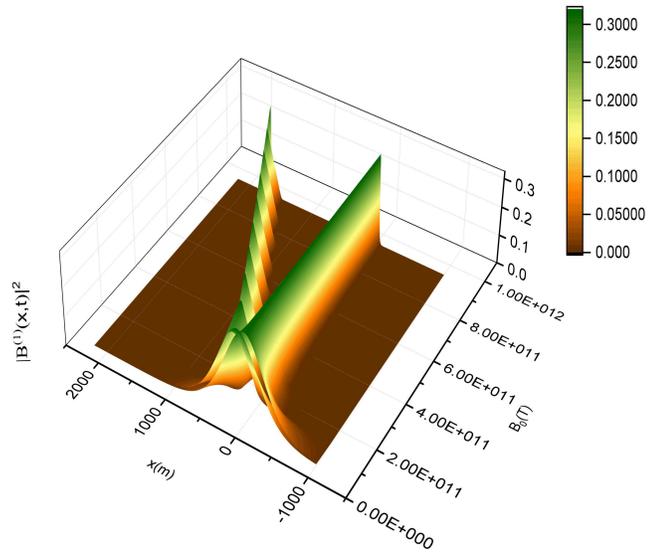}
\end{tabular}}
 \caption{\footnotesize
 Soliton profiles of the transverse magnetic field perturbation at $ t = 0 s $ and $ t =  10^{-3} s $ for different values of background magnetic field $ | B_0 | $ by considering  $\rho_{0}=3\times 10^{17} kg/m^3$. }
\end{figure}

Figure 4 demonstrates soliton profiles created in media with different values of mass density. This figure shows that the width of solitary wave increases as the mass density is given rise. But the wave phase speed behaves differently and it decreases with increasing values of the mass dinsity. Although the width of the soliton is not very sensitive to the changes of mass density.

\begin{figure}[htp]\label{fig4}
\centerline{\begin{tabular}{cc}
\includegraphics[width=10 cm, height= 10 cm]{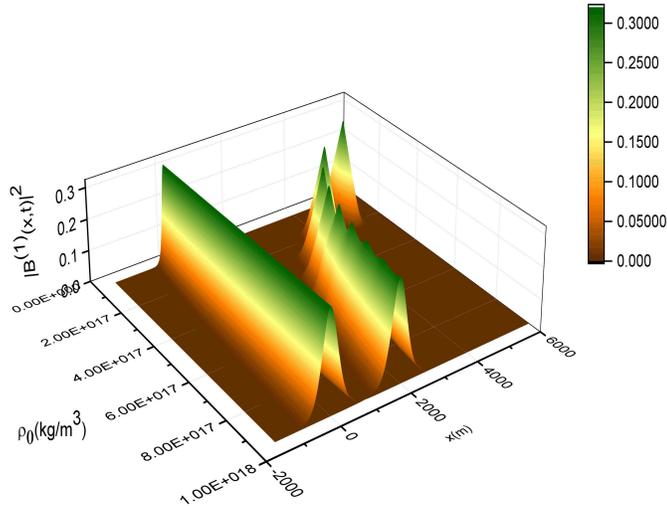}
\end{tabular}}
 \caption{\footnotesize
 Soliton profiles of the transverse magnetic field perturbation at $ t = 0 s $ and $ t = 2\times 10^{-2} s $ for different values of background mass density $  \rho_0  $ by considering  $ |B_0|= 10^{11} T$. }
\end{figure}

\section*{7. Conclusions and remarks }
We have presented the magnetohydrodynamic equations in a non relativistic framework for cold QGP environments. Considering the cold QGP as a two-fluid MHD model ($ u , d $ quarks) and combining the hydrodynamic equations of the system with the
equation of state, according to the MIT bag model, eventuates a complete set of equations with nonlinear and dispersive terms which play essential roles in the dynamics of the system. Time evolution of small amplitude perturbations in density and/or magnetic field  have been investigated by applying the reductive perturbation method (RPM). The result is the modified derivative nonlinear Schrodinger (mDNLS) equation which governs the mass density (localized transverse magnetic field) perturbation waves. These solitary waves can be propagated without any distortion in its initial direction in cold QGP. The width and the velocity of these localized waves are functions of the background mass (or baryon) density and magnetic field.  The solitonic wave phase speed increases as the magnetic field $ B_0 $ increases, while it decreases by increasing values of the mass density $ \rho_0 $. Also, increasing the magnetic field $ B_0 $ reduces the width of solitary waves; but variation of the mass density has not significant effects in the characteristics of solitary waves. Therefore, we can conclude that the solitary profiles are expected to be established in such media when the background magnetic field is in a specific range.

The most important result of this work is that, the small amplitude propagation of localized waves in cold QGP is not governed by the KdV equation. Indeed such waves are solitons of the modified DNLS equation which behave very different from the KdV localized solutions.  Derived equations also clearly indicates that solitary waves in cold QGP (by considering its electromagnetic effects) are completely stable, independent from the existence of Laplacian terms in the energy density (which are generally weak).\

Similar investigation can be done for neutron stars using proper EOS like \cite{ne}. It is expected that localized perturbation in magnetic field and energy density also appears in this media. In other word, it is expected that presented behaviour may be observed in different structures of compact astrophysical objects.

So there are many works in this subject which should be done (or revised). It is possible that there exist some sorts of instabilities in the propagating waves as solutions of DNLS equations, which should be investigated. Same problem but with different models for the equation of state can be solved, and results should be compared. The problem also is open for other forms of super dense media, like hadronic gas and nuclear matter as expected to be found in compact astrophysical objects. 

\section*{acknowledgement}
This work is supported by the Ferdowsi University of Mashhad under the Grant NO. $ 3/28787 $

\end{document}